\documentstyle[preprint,aps,prb,epsfig]{revtex}
\begin{document}
\draft

\preprint{}

\title{Variational bounds on the ground-state energy of three electrons and one hole in two-dimension}
\author{Nie Luo}
\address{Department of Physics and Astronomy, Northwestern University, Evanston, IL 60208}
\date{\today}
\maketitle
\begin{abstract} 

We consider a model of three electrons and one hole confined in a two-dimensional (2D) plane, interacting with one another through Coulomb forces. Using a Ritz variational method  we find an upper bound of $\approx -0.0112me^4/8\pi^2 \epsilon ^2 \hbar ^2$ for the ground-state energy of such a system when the particles are near one another. The possible connections of such a complex to other fields of physics are discussed. 

\end{abstract}
%

\section{Introduction}

The research on trions in 2D systems has shown many interesting results in recent years\cite{kheng,huard,stebe1,thilagam}. In a broader sense, the trion problem, not limited to 2D, is linked to the existence and the stability of negative hydrogen ion H$^-$ which has not only important implication for atomic physics and astrophysics\cite{bethe}, but also vital application to other fields such as physical chemistry\cite{massey}, accelerator engineering\cite{acl}, and controlled fusion\cite{fusion}. 

In a trion, or charged exciton, two electrons are bound to one hole despite the strong Coulomb repulsion between electrons. (Likewise two holes can be bound to one electron as well but the discussion that follows is the same if ``hole'' is switched with ``electron.'') It is quite a marvel that the two electrons would not fall apart from the strong Coulomb repulsion between them. More surprising is that the ground-state energy of such a complex is {\it lower} than that of an exciton, at least in the limit of heavy hole ({\it i.e.}, the mass of the hole is much larger than that of electrons, which effectively results into the H$^-$ problem). 

First let us try to understand in terms of classical mechanics why trion forms despite the strong Coulomb repulsion. 

\begin{figure} [t]
\begin{center}
\epsfxsize=3.4in 
\epsfysize=0.8in 
\epsffile{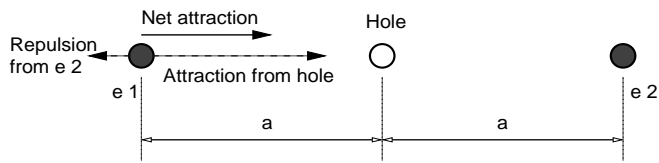}
\end{center}
\caption{Two electrons and a hole arrange themselves to form the bound state of trion in a classical picture of electrodynamics. Arrows indicate forces. Electrons e1 and e2 are shown as the shaded circles.}
\label{fig1}      
\end{figure} 

Suppose that two electrons are on two sides of one hole as shown in Fig. 1. The electrostatic forces felt by electron 1 (e1) are from the hole and electron 2 (e2). Because the forces are governed by the inverse square Coulomb's law, the repulsion exerted on e1 by e2 is only $1/4$ that of the attraction given by the hole. Similar argument applies to e2 and therefore these two electrons tend to come together to form a bound state in spite of the strong repulsion among like charges. In order for electrons not to fall into the hole we can assume in a classical fashion that the electrons rotate around the hole in such a way that the inward electrostatic force is balanced by the centrifugal force. In quantum mechanics such tendency of electron's coming together will be stopped by the penalty in kinetic energy which increases as $a^{-2}$, where $a$ is roughly the average distance between the hole and one of the electrons. Note that the gain in Coulomb energy varies at a slower pace approximately as $a^{-1}$ so that it would be hard for the electrons to come near the hole without limit, if the possible electron-hole recombination is not taken into account. (The trion will not be stable if the recombination is spontaneous. We will discuss this problem further in Sec. \ref{stbl-eh-rcb}.)

Therefore the bound state of trion is not quite a surprise after all. It basically results from the inverse square Coulomb's law and a favorable geometric configuration of two electrons and one hole.

\section{Three electrons and one hole}

Now we want to take one step further to see if a similar physical picture generalizes to the case of three electrons and one hole. In the heavy hole limit, this problem is in effect that for the doubly charged negative hydrogen ion H$^{--}$. There is a much celebrated theorem\cite{lieb1,lieb2} by E. H. Lieb saying that  H$^{--}$ is {\it not} stable against dissociation into  H$^{-}$  and e$^{-}$ in {\it vacuum}. Here vacuum means that there is not any other charged particle nearby at finite distance except for the particles forming the H$^{--}$ ion. We need a few more words on this theorem. In Lieb's original papers, a bound system is defined as an eigenstate of the Hamiltonian. This is a strong condition, which leaves room for metastable states for H$^{--}$. The existence of possible metastable states for H$^{--}$ is currently a highly debated issue\cite{h2-0,h2-1,h2-2,h2-3,h2-4}. Lieb's theorem either does not rule out any possible H$^{--}$ state whose total energy is negative. Here the energy zero is taken when particles are infinitely separated and thus with arbitrarily small kinetic and potential energy.  

In this paper we study the bound on the ground-state energy of three electrons and one hole at {\it close} distance in 2D. Instead of having a heavy positive charge, we assume that the masses of the four are all the same. Our result shows strictly that the ground-state energy is negative when the four particles are relatively localized one to another.  

\begin{figure} [t]
\begin{center}
\epsfxsize=3.4in 
\epsfysize=3.6in 
\epsffile{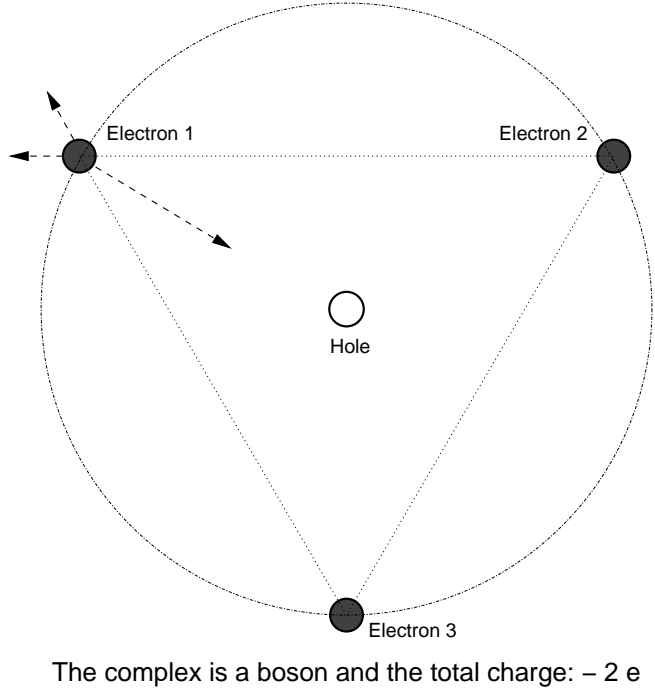}
\end{center}
\caption{Three electrons and a hole arrange themselves to form a ``bound'' state in a classical picture. Arrows indicate forces. The dotted equilateral triangle is a guide to the eye, showing three electrons at equal distance. The dash-dotted circle traces the orbit of electrons.}
\label{fig2}      
\end{figure} 

Before starting quantum mechanical calculation, we want to show in a classical picture why such a state of negative energy might be possible. In Fig. 2, three electrons are at the vertices of an equilateral triangle, with one hole in the center. According to the simple knowledge of electrostatics, the force on the central hole is 0 while those on the electrons are $\frac{(\sqrt{3}-1)e^2}{4\sqrt{3}\pi \epsilon a^2}$ inwards to the center, where $a$ is the distance between the hole and one electron. Not to make themselves fall into the hole the electrons may rotate around the hole in such a way that the inward electrostatic force is balanced by the centrifugal force, just as in a trion. Intuitively a four-body system with a net charge of $-2e$ may not be stable because of the strong Coulomb repulsion. However under the special geometry as shown, the peculiarity of Coulomb's law (an inverse square law) actually favors a bound state of 3 electrons and 1 hole (3E+1H), at least within the framework of classical electrodynamics. Thus these electrons tend to move to the hole, giving rise to an effective binding among them. 

Note that the total charge of such a complex is $-2e$ which just reminds us of the Cooper pair that brings in superconductivity, because the complex of four fermions (three electrons and one hole) is now effectively a charged boson. 

The stability of such complex against electron-hole recombination or annihilation actually depends on the energy band structure of the host conductors. While  electrons and holes recombine spontaneously in semiconductors, they coexist peacefully inside semimetals without destroying one another. The stability is governed by semi-metallic energy band structures. We will deliberate on this later in Sec. \ref{stbl-eh-rcb}.

For the purpose of convenience, let us introduce a new term, {\it tetron}, to refer to the 3E+1H complex, in a sense similar to the term trion. We will use tetron and 3E+1H interchangeably thereafter. 

Our strategy for this paper is as follows. We will first mention relevant 2D quantum mechanical results for the Coulomb potential in Sec. \ref{qm2d}. Then as a warm-up exercise we try to find out the upper bound on the ground-state energy of a 2D trion in the heavy hole approximation, which is given in Sec. \ref{sec-trion}. Based on techniques developed for trions, we try a few fully antisymmetrized wavefunctions in Sec. \ref{sec-tetron} to show that states of negative energy indeed happen to tetron.  

\section{Quantum mechanics of an electron in 2D Coulomb potential} \label{qm2d}

This problem has been worked out numerous times but we just repeat some relevant results here for quick reference. 

The Hamiltonian for an electron in a 2D Coulomb potential reads 
\begin{equation}
H = - \frac{\hbar^2}{2m} \nabla ^2 - \frac{e^2}{4\pi \epsilon r},
\end{equation}
where $\epsilon$ is the permittivity, $e$ the elementary charge and $m$ the mass of the electron. 

We try to solve the problem in a polar coordinate, where the gradient operator is given by 
\begin{equation}
\nabla = {\bf {\hat r}}\frac{\partial}{\partial r} + \bbox{{\hat \theta}} \frac{1}{r} \frac{\partial}{\partial \theta},
\end{equation}  
with the corresponding Laplacian
\begin{equation}
\nabla ^2 = \frac{1}{r}\frac{\partial}{\partial r}(r\frac{\partial}{\partial r}) +  \frac{1}{r^2} \frac{\partial^2}{\partial \theta ^2}.
\end{equation}  

The Schr\"{o}dinger equation reads 

\begin{equation}
 \{ - \frac{\hbar^2}{2m} \nabla ^2 - \frac{e^2}{4\pi \epsilon r} \} \Psi = E \Psi .
\label{Schroedinger} 
\end{equation}

We want to simplify Eq. (\ref{Schroedinger}) by making it dimensionless.  
Note that $\hbar^2/2m$ has a dimension of (energy)(length)$^2$ while $e^2/4\pi \epsilon $ is in (energy)(length). We select the units (or the energy and length scales) so that both $\hbar^2/2m$ and  $e^2/4\pi \epsilon $ are 1. Under such a convention length will be measured in the unit $2\pi \epsilon\hbar^2/me^2$ while energy is measured in $me^4/8\pi^2 \epsilon^2 \hbar^2$. This unit system is summarized in Table \ref{table1}. 
\begin{table}
\caption{The units chosen to make the Schr\"{o}dinger equation dimensionless}
\begin{tabular}{cccccc}
Dimension &(energy)(length)$^2$&(energy)(length)&(length)&(energy)&(wavenumber)\\
\tableline
Ordinary unit &$\frac{\hbar^2}{2m}$&$ \frac{e^2}{4\pi \epsilon }$&$\frac{2\pi \epsilon\hbar^2}{me^2}$&$\frac{me^4}{8\pi^2 \epsilon^2 \hbar^2}$&$\frac{me^2}{2\pi \epsilon\hbar^2}$\\
Dimensionless unit &1&1&1&1&1\\
\end{tabular}
\label{table1}
\end{table}

Now the Schr\"{o}dinger equation takes a simple form,
\begin{equation}
 \{ - \nabla ^2 - \frac{1}{r} \} \Psi = E \Psi. 
\label{Schroedinger-dmless} 
\end{equation}

Eq. (\ref{Schroedinger-dmless}) has the following solutions as well known,

\begin{equation}
\Psi_{nm}(r,\theta) = \frac{w_{nm}(r)}{\sqrt{r}}\frac{\exp(\pm im\theta)}{\sqrt{2\pi}}, m=0, 1, 2, 3,\dots ,
\end{equation}
where $w_{nm}(r)$ is given by
\begin{eqnarray}
w_{nm}(r) &=& \frac{(2Kr)^{m+1/2}\exp(-Kr)}{(2m)!}[\frac{K(n+m-1/2)!}{n(n-m-1/2)!}]^{1/2} \nonumber \\
&&\times _{1}F_{1}(-n+m+\frac{1}{2}; 2m+1; 2Kr).  
\end{eqnarray}
Here $_{1}F_{1}$ is the confluent hypergeometric function, and
\begin{equation}
K=\frac{1}{2n}, n=\frac{1}{2}, \frac{3}{2},\dots , m= 0, 1,\dots, n-\frac{1}{2}.
\end{equation}

The energy $E_{nm}$ of eigenstate $\Psi_{nm}$ is given by 
\begin{equation}
E_{nm}= -\frac{1}{4n^2}.
\end{equation}

For the purpose of convenience, we list the first four eigenstates as follows,
\begin{eqnarray}
 \Psi_{\frac{1}{2}, 0}(r,\theta)&=& \sqrt{\frac{2}{\pi}} \exp(-r), \nonumber \\
 \Psi_{\frac{3}{2}, 0}(r,\theta)&=& \sqrt{\frac{2}{27\pi}} [1-\frac{2}{3}r] \exp(-\frac{r}{3}), \nonumber \\
 \Psi_{\frac{3}{2}, \pm 1}(r,\theta)&=& \sqrt{\frac{4}{243\pi}} r \exp(-\frac{r}{3}) \exp(\pm i \theta). \nonumber \\ \label{3psi}
\end{eqnarray} 

\section{Upper bound on  ground-state energy of a 2D trion}\label{sec-trion}

We take this problem as a warm-up exercise. This is a very illustrative example showing how to balance between the Coulomb and kinetic energy. It also gives us some intuition in choosing good trial wavefunctions and on how to deal with two-body Coulomb integrals. The basic line of reasoning follows that of H.A. Bethe\cite{bethe}, E.A. Hylleraas\cite{trion-more-stbl1} and others\cite{trion-more-stbl2}. Although there have been recent researches\cite{stebe1,thilagam} in the same direction, our methods in dealing with the integrals are somehow different and hopefully easier to understand.  

For the purpose of simplicity we take the mass of the hole as infinity. The Schr\"{o}dinger for this problem is now clearly

\begin{equation}
 \{ - \nabla _1 ^2 - \nabla _2 ^2 - \frac{1}{r_1} - \frac{1}{r_2} + \frac{1}{r_{12}} \} \Psi = E \Psi. 
\label{Schroedinger-dmless-trion} 
\end{equation}
where $\nabla _1 ^2$ and $\nabla _2 ^2$ are respectively short-hands for 
\begin{equation}
\frac{1}{r_1}\frac{\partial}{\partial r_1}(r_1\frac{\partial}{\partial r_1}) +  \frac{1}{r_1^2} \frac{\partial^2}{\partial \theta_1 ^2},
\end{equation}
and 
\begin{equation}
\frac{1}{r_2}\frac{\partial}{\partial r_2}(r_2\frac{\partial}{\partial r_2}) +  \frac{1}{r_2^2} \frac{\partial^2}{\partial \theta_2 ^2}.
\end{equation}
Here subscripts refer to electrons such that $r_1$,  $\theta_1$ and $r_2$, $\theta_2$ are the coordinates of e1 and e2 respectively. 

We choose the following trial wavefunction  
\begin{equation}
\Psi = \exp [-\lambda (r_1 + r_2)][1+k(r_1 -r_2)^2], \label{14}
\end{equation}
where $\lambda$ and $k$ are parameters to be determined later by variational principle. The first term in this wavefunction suggests that e1 and e2 are both in $1s$ states. Here $1s$ state are not exactly the $\Psi_{\frac{1}{2}, 0}$ given in Eqs. (\ref{3psi}) but rather that with a screening constant $\lambda$. The physical origin of $\lambda$ is clear: the Coulomb field of the hole felt by one electron is reduced (screened) by the presence of the other. For this reason we also require that $\lambda$ be positive and smaller than 1. The existence of the second term is to reduce the Coulomb interaction between e1 and e2. Note that this term is small when $r_1$ and $r_2$ are close which is another way of saying that electrons 1 and 2 tend to avoid each other. So why not just keep the second term only? The reason is that while this term is good for reducing the Coulomb energy it causes some penalty in kinetic energy when compared with the first one. Thus the viable solution is to combine the merits of the two terms, which results in the trial wavefunction (\ref{14}). 

To facilitate our calculation we want better management on the trial wavefunction. Note that functions like $\exp[-\lambda (r_1 + r_2)]$ are not normalized and would be a headache for the algebra and calculus to come. The solution is to generalize the eigenstates of Eqs. (\ref{3psi}) to accommodate the screening effect. The first four solutions now read  
\begin{eqnarray}
 |1s \rangle &=& \sqrt{\frac{2}{\pi}} \lambda \exp(-\lambda r),  \\
 |2s \rangle &=& \sqrt{\frac{2}{27\pi}} [\lambda -\frac{2}{3}\lambda ^2 r] \exp(-\frac{\lambda r}{3}), \\
 |2p^{\pm} \rangle &=& \sqrt{\frac{4}{243\pi}}  \lambda ^2  r \exp(-\frac{\lambda r}{3}) \exp(\pm i \theta). 
\end{eqnarray}
where we have used notations $1s$, $2s$, and $2p^{\pm}$ in place of the cumbersome $\Psi$. The meaning of these symbols should be self-evident given a background knowledge of hydrogen eigenstates in three-dimension (3D). It is not hard to verify that these wavefunctions are now normalized.

The trial wavefunction is then instead written as
\begin{equation}
\Psi = [1+k(r_1 - r_2)^2]|1s_1 1s_2\rangle,
\end{equation}
where $| 1s_1 1s_2\rangle = | 1s_1\rangle |1s_2\rangle  $ stands for 
\begin{equation}
 \sqrt{\frac{2}{\pi}} \lambda \exp(-\lambda r_1)  \sqrt{\frac{2}{\pi}} \lambda \exp(-\lambda r_2),
\end{equation} 
and $| 1s_1\rangle$ is the short-hand for 
\begin{equation}
 \sqrt{\frac{2}{\pi}} \lambda \exp(-\lambda r_1). 
\end{equation} 
Similarly we define $| 1s_2 \rangle$.

This trial wave function is symmetric in the spatial coordinates and in order to make the total wavefunction antisymmetric we only need to multiply $\Psi$ by a spin singlet state ${\cal S}_{12}^a = \frac{1}{\sqrt{2}}[ |\uparrow _1 \downarrow _2 \rangle -  |\downarrow _1 \uparrow _2 \rangle ]$ where the subscripts are electron indices with the superscript $a$ indicating antisymmetry. Note that the spin wavefunction in this case will not affect the calculation of either $\langle \Psi {\cal S}_{12}^a | \Psi {\cal S}_{12}^a \rangle = \langle \Psi | \Psi \rangle  $  or $\langle \Psi {\cal S}_{12}^a |H| \Psi {\cal S}_{12}^a \rangle = \langle \Psi  |H| \Psi \rangle$ thus we can drop it for now. We also tabulate some often used integrals in Table \ref{table2} to facilitate our calculation. 

\begin{table}
\caption{Some useful integrals involving $|{\rm 1s} \rangle$ state}
\begin{tabular}{cccccc}
Integral&$\langle {\rm 1s} |r^{-1}|{\rm 1s} \rangle$&$\langle {\rm 1s} |r|{\rm 1s} \rangle$&$\langle {\rm 1s} |r^2|{\rm 1s} \rangle$&$\langle {\rm 1s} |r^3|{\rm 1s} \rangle$&$\langle {\rm 1s} |r^4|{\rm 1s} \rangle$ \\
\tableline
Value&$2\lambda$&$\lambda^{-1}$&$\frac{3}{2}\lambda^{-2}$&$3\lambda^{-3}$&$\frac{15}{2}\lambda^{-4}$\\
\end{tabular}
\label{table2}
\end{table}

The Ritz variational principle states
\begin{equation}
E(\Psi)= \frac{\langle \Psi |H| \Psi \rangle}{\langle \Psi | \Psi \rangle} = {\rm min.},
\end{equation}
where 
\begin{equation}
\langle \Psi |H| \Psi \rangle = \langle \Psi |  - \nabla _1 ^2 - \nabla _2 ^2 - \frac{1}{r_1} - \frac{1}{r_2} + \frac{1}{r_{12}} | \Psi \rangle. \label{psiHpsi-trion}
\end{equation}

First we need to calculate $ \langle \Psi | \Psi \rangle  $, 
\begin{eqnarray}
\langle \Psi | \Psi \rangle &=&  \langle 1s_1 1s_2 | \{ 1+  2k(r_1^2+r_2^2 -2r_1r_2) \nonumber \\ 
&&+ k^2[r_1^4 + r_2^4 + 6r_1^2r_2^2-4r_1^3r_2-4r_1r_2^3] \} |1s_1 1s_2 \rangle \nonumber \\
&=& 1+ 2k\lambda^{-2} +4.5k^2\lambda^{-4}.
\end{eqnarray}

The first term on the right-hand-side (RHS) of Eq. (\ref{psiHpsi-trion}) is 
\begin{equation}
\langle \Psi |- \nabla_1^2| \Psi \rangle = \langle \nabla_1 \Psi |\nabla_1 \Psi \rangle,
\end{equation}
where $ |\nabla_1 \Psi \rangle $ is to be understood as $\nabla_1 |\Psi \rangle$, while $ \langle \nabla_1 \Psi | $, as a notation, is the Hermitian conjugate of  $ |\nabla_1 \Psi \rangle $.  This identity can be proved in rectangular Cartesian coordinate using integration by parts when $\Psi$ is zero at infinity which our trial wavefunction satisfies. 

Because 
\begin{eqnarray}
\nabla_1 | \Psi \rangle &=& \{ {\bf {\hat r}}_1 \frac{\partial}{\partial r_1} + {\bbox{\hat \theta}_1} \frac{1}{r_1} \frac{\partial}{\partial \theta_1} \} | 1s_1 1s_2 \rangle \nonumber \\ 
&=& \{ -\lambda [1+k(r_1^2+r_2^2 -2r_1r_2)] + 2k(r_1- r_2) \} {\bf {\hat r}}_1 | 1s_1 1s_2 \rangle  \nonumber \\ 
&=& - \lambda \Psi {\bf {\hat r}}_1 +  2k(r_1 - r_2) {\bf {\hat r}}_1 | 1s_1 1s_2 \rangle,
\end{eqnarray}
we have 
\begin{eqnarray}
\langle \nabla_1 \Psi |\nabla_1 \Psi \rangle &=& \lambda ^2 \langle \Psi | \Psi \rangle  + 4k^2(3\lambda^{-2} -2\lambda^{-2} )  \nonumber \\ 
&=& \lambda ^2 + 2k + 8.5 k^2 \lambda ^{-2}. \label{26}
\end{eqnarray} 

Since $\Psi$ is symmetric in $r_1$ and $r_2$ it should not be a surprise that the second term on RHS of Eq. \ref{psiHpsi-trion},  $\langle \Psi |- \nabla_2^2| \Psi \rangle$ is equal to $\langle \Psi |- \nabla_1^2| \Psi \rangle $. 

The third term in  $ \langle \Psi |H| \Psi \rangle  $ [{\it i.e.}, Eq. (\ref{psiHpsi-trion})] is  
\begin{eqnarray}
-\langle \Psi |\frac{1}{r_1} | \Psi \rangle &=&  -\langle 1s_1 1s_2 |\frac{1}{r_1} \{ 1+  2k(r_1^2+r_2^2 -2r_1r_2) \nonumber \\ 
&&+ k^2[r_1^4 + r_2^4 + 6r_1^2r_2^2-4r_1^3r_2-4r_1r_2^3] \} |1s_1 1s_2 \rangle \nonumber \\
&=& -(2\lambda+4k\lambda^{-1}+ 9k^2\lambda^{-3})  \label{27}
\end{eqnarray} 
to which the fourth term $-\langle \Psi |\frac{1}{r_2}| \Psi \rangle$ is equal because of the symmetry in $r_1$ and $r_2$. 

The last term of  $\langle \Psi |H| \Psi \rangle$ is a two-body Coulomb integral involving elliptic integrals, and
\begin{eqnarray}
\langle \Psi |\frac{1}{r_{12}} | \Psi \rangle &=&  \langle 1s_1 1s_2 |\frac{1}{r_{12}} \{ 1+  2k(r_1^2+r_2^2 -2r_1r_2) \nonumber \\ 
&&+ k^2[r_1^4 + r_2^4 + 6r_1^2r_2^2-4r_1^3r_2-4r_1r_2^3] \} |1s_1 1s_2 \rangle \nonumber \\
&=& \langle {  1s_1 1s_2} |\frac{1}{r_{12}}\{ 1+  2k(2r_1^2-2r_1r_2) \nonumber \\ 
&&+ k^2[2 r_1^4 + 6r_1^2r_2^2-8r_1^3r_2] \} |{  1s_1 1s_2} \rangle,  \label{trion-two-body}
\end{eqnarray} 
where we have used identities 
\begin{eqnarray}
\langle {  1s_1 1s_2} |\frac{r_2^2}{r_{12}} |{  1s_1 1s_2} \rangle &=& \langle {  1s_1 1s_2} |\frac{r_1^2}{r_{12}} |{  1s_1 1s_2} \rangle,\nonumber \\
\langle {  1s_1 1s_2} |\frac{r_2^4}{r_{12}} |{  1s_1 1s_2} \rangle &=& \langle {  1s_1 1s_2} |\frac{r_1^4}{r_{12}} |{  1s_1 1s_2} \rangle, \nonumber \\
\langle {  1s_1 1s_2} |\frac{r_1r_2^3}{r_{12}} |{  1s_1 1s_2} \rangle &=& \langle {  1s_1 1s_2} |\frac{r_1^3r_2}{r_{12}} |{  1s_1 1s_2} \rangle,  \nonumber \\
\end{eqnarray}
which are easy to understand because of the symmetry of these integrals in $r_1$ and $r_2$. 

As an example, we explicitly show in the following how to work out the first term on RHS of Eq. (\ref{trion-two-body}).
\begin{eqnarray}
&&\langle {  1s_1 1s_2}|\frac{1}{r_{12}}|{  1s_1 1s_2}\rangle \nonumber \\
&&=(\frac{2}{\pi})^2 \lambda^4 \int_0^\infty r_1 dr_1  \int_0^\infty r_2 dr_2 \exp[-2\lambda(r_1+r_2)]  \int_0^{2\pi} d\theta_1 \int_0^{2\pi} d\theta_2 \frac{1}{\sqrt{r_1^2+r_2^2-2r_1r_2\cos\theta_{12}}}, \label{int12}
\end{eqnarray} 
where $\theta_{12} = \theta_1 -\theta_2$. 

We want to change the integration variables in the integral 
\begin{equation}
\int_0^{2\pi} d\theta_1 \int_0^{2\pi} d\theta_2 \frac{1}{\sqrt{r_1^2+r_2^2-2r_1r_2\cos \theta_{12}}} \label{theta12}
\end{equation}
from $\theta_1$ and $\theta_2$ to $\theta_{12}$ and $\theta_2$. To understand how this is done, we refer to Fig. 3. Shown in Fig. 3 is a plane formed by $\theta_1$ and $\theta_2$. The integration limits of $\theta_1$ and $\theta_2$ in Eq. \ref{theta12}  or the integration domain corresponds to the square ABCD.

\begin{figure} [h]
\begin{center}
\leavevmode
\epsfxsize=3.1in 
\epsfysize=3.5in 
\epsffile{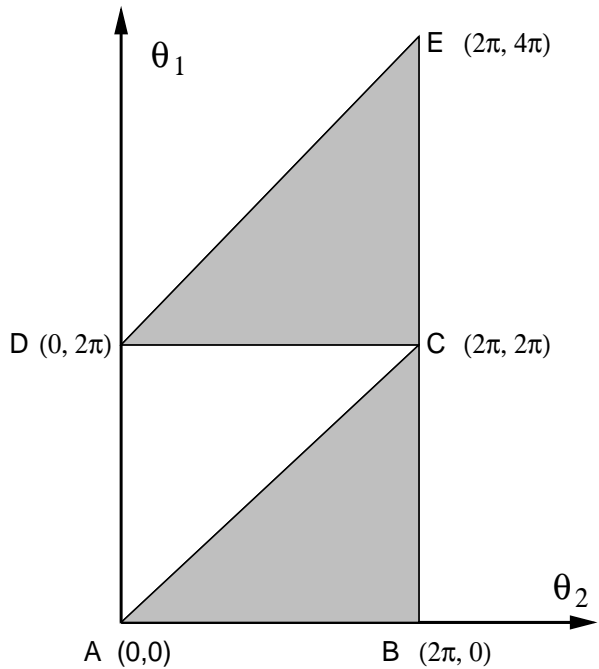}
\end{center}
\caption{Change of variable from $\theta_1$  to $\theta_{12}$ in integral (31).}
\label{fig3}      
\end{figure}

Because the integrand in integral (\ref{theta12})
\begin{equation}
\frac{1}{\sqrt{r_1^2+r_2^2-2r_1r_2\cos\theta_{12}}}
\end{equation}
is periodic in $\theta_1$ with period $2\pi$, the integration over triangle ABC is the same as that over triangle DCE. Note that DCE is the translate of ABC by $2\pi$ in the $\theta_1$ direction. Thus we can now change the domain of integration from the square ABCD to the parallelogram ACED. Since the segments AC and DE respectively satisfy the relation $\theta_1 - \theta_2 = \theta_{12}= 0$  and $\theta_1 - \theta_2 = \theta_{12} = 2 \pi$, if we transform integral variables to $\theta_{12}$ and $\theta _2$ along with the change in integration domain, we will have 
\begin{eqnarray}
&&\int_0^{2\pi} d\theta_1 \int_0^{2\pi} d\theta_2 \frac{1}{\sqrt{r_1^2+r_2^2-2r_1r_2\cos\theta_{12}}} \nonumber  \\
&=& \int_0^{2\pi} d\theta_2 \int_0^{2\pi} d\theta_{12} \frac{1}{\sqrt{r_1^2+r_2^2-2r_1r_2\cos\theta_{12}}} | \frac{\partial (\theta_1, \theta_2)}{\partial (\theta_{2}, \theta_{12})}|,
\end{eqnarray}   
where $|\frac{\partial (\theta_1, \theta_2)}{\partial (\theta_{2}, \theta_{12})}|$ is the absolute of Jacobian determinant
\begin{equation}
\frac{\partial (\theta_1, \theta_2)}{\partial (\theta_{2}, \theta_{12})} =
\left| 
\begin{array}{cc}
\frac{\partial \theta_1}{\partial \theta_{2}} & \frac{\partial \theta_1}{\partial \theta_{12}} \\
\frac{\partial \theta_2}{\partial \theta_{2}} & \frac{\partial \theta_2}{\partial \theta_{12}} \\
\end{array}
\right|=
\left|
\begin{array}{cc}
 1 & 1 \\
 1 & 0 \\
\end{array}
\right|=-1,
\end{equation}
which is identically $1$ in its absolute value.
 
Thus integral (\ref{theta12}) now satisfies 
\begin{eqnarray}
&&\int_0^{2\pi} d\theta_2 \int_0^{2\pi} d\theta_{12} \frac{1}{\sqrt{r_1^2+r_2^2-2r_1r_2\cos\theta_{12}}} \nonumber \\
&=& 2 \pi \int_0^{2\pi} d\theta_{12} \frac{1}{\sqrt{r_1^2+r_2^2-2r_1r_2\cos\theta_{12}}}. \label{35}
\end{eqnarray}

Inserting Eq. (\ref{35}) into Eq. (\ref{int12}), we get
\begin{eqnarray}
&&\langle {  1s_1 1s_2}|\frac{1}{r_{12}}|{  1s_1 1s_2}\rangle \nonumber \\
&&=\frac{8}{\pi} \lambda^4 \int_0^\infty dr_1  \int_0^\infty dr_2 \exp[2\lambda(r_1+r_2)]r_1 r_2 \int_0^{2\pi} d\theta_{12} \frac{1}{\sqrt{r_1^2+r_2^2-2r_1r_2\cos\theta_{12}}}. \label{int12-simple}
\end{eqnarray}

Eq. (\ref{int12-simple}) can be further simplified by the following change of variables,
\begin{eqnarray}
&&r_1=r, \nonumber \\ 
&&r_2=ur,
\end{eqnarray}
with corresponding Jacobian 
\begin{equation}
\frac{\partial (r_1, r_2)}{\partial (r, u)} =
\left| 
\begin{array}{cc}
\frac{\partial r_1}{\partial r} & \frac{\partial r_1}{\partial u} \\
\frac{\partial r_2}{\partial r} & \frac{\partial r_2}{\partial u} \\
\end{array}
\right| = 
\left| 
\begin{array}{cc}
 1 & 0 \\
 u & r \\
\end{array}
\right|
=r.
\end{equation}

This transformation changes RHS of Eq. (\ref{int12-simple}) to  
\begin{equation}
\frac{8}{\pi} \lambda^4 \int_0^\infty u du \int_0^{2\pi} d\theta_{12} \frac{1}{\sqrt{1+u^2-2u\cos\theta_{12}}} \int_0^\infty dr r^2 \exp[2\lambda(1+u)r], \label{int12-simple-ru}
\end{equation}
which can be first integrated with respect to $r$ to give 
\begin{equation}
\frac{8}{\pi} \lambda^4 2! \int_0^\infty [\frac{2}{\lambda(1+u)}]^3 u du \int_0^{2\pi} d\theta_{12} \frac{1}{\sqrt{1+u^2-2u\cos\theta_{12}}}, \label{int12-simple-u}
\end{equation}
where 
\begin{equation}
\int_0^{2\pi} d\theta_{12} \frac{u}{\sqrt{1+u^2-2u\cos\theta_{12}}} =
\frac{4\bbox{K}(\frac{-4u}{1-2u+u^2})}{\sqrt{1-2u+u^2}}
\end{equation}
after integration over $\theta_{12}$, with $\bbox{K}$ the complete elliptic integral of the first kind. Integral (\ref{int12-simple-u}) is therefore reduced to 
\begin{equation}
\frac{16}{\pi} \lambda  \int_0^\infty [\frac{2}{1+u}]^3 \frac{4u\bbox{K}(\frac{-4u}{1-2u+u^2})}{\sqrt{1-2u+u^2}}  du
\end{equation}
which can be integrated numerically to give the final answer 
\begin{eqnarray}
\langle {  1s_1 1s_2}|\frac{1}{r_{12}}|{  1s_1 1s_2}\rangle = 1.1780972 \lambda.\label{43}
\end{eqnarray} 

Similarly we work out other terms in Eq. (\ref{trion-two-body})
\begin{eqnarray}
& &2k\langle {  1s_1 1s_2}|\frac{2r_1^2-2r_1r_2}{r_{12}}|{  1s_1 1s_2}\rangle = 1.0308351 k \lambda^{-1},\label{44}
\end{eqnarray}
and
\begin{eqnarray}
& &k^2 \langle {  1s_1 1s_2}|\frac{2r_1^4+6r_1^2r_2^2-8r_1^3r_2}{r_{12}}|{  1s_1 1s_2}\rangle \nonumber \\
&& = 1.5186410 k^2 \lambda^{-3}. \label{45}
\end{eqnarray} 

Substituting Eqs. (\ref{43}),(\ref{44}) and (\ref{45}) into (\ref{trion-two-body}) and combining terms from (\ref{26}), (\ref{27}), we find the expectation value of the Hamiltonian 
\begin{eqnarray}
&&\langle \Psi |H| \Psi \rangle \nonumber \\
&&= 2 \lambda ^2 + 4 k + 17 k^2 \lambda^{-2} \nonumber \\
&&-2.8219028 \lambda - 6.9691649 k \lambda^ {-1} -16.4813590 k^2 \lambda^{-3},
\end{eqnarray}
where the first three terms are from kinetic energy and the last three are Coulomb ones. 

We want to minimize the energy
\begin{eqnarray}
E &=&\frac {\langle \Psi |H| \Psi \rangle}{\langle \Psi|\Psi\rangle} \nonumber \\
&=& \frac{ 2 \lambda ^2 + 4 k + 17 k^2 \lambda^{-2} -2.8219028 \lambda - 6.9691649 k \lambda^ {-1} -16.4813590 k^2 \lambda^{-3}}{1+2k\lambda^{-2} +4.5k^2\lambda^{-4}},
\end{eqnarray}
with respect to $\lambda$ and $k$, which has numerical results as follows,
\begin{equation}
E_{min} = -1.08147; k=0.111270; \lambda = 0.693001. 
\end{equation} 

This value, as an upper bound of the ground-state energy for our specific trion, is lower than that of an exciton with heavy hole, which is $-1$. The trion is therefore more stable than exciton and it forms spontaneously given one exciton and an extra electron. 

Our calculation shows that the trion binding energy in two-dimension $> 0.08147$ 2D Rydberg, larger than that of 3D one which is $\approx 0.055$ 3D Rydberg. This is not a surprise because it is well known that lower dimensionality favors the formation of bound states. As a result, pure excitons in 2D have a binding energy 4 times as big as that in 3D.

We are so far confined to heavy hole limit. It is interesting to see quite some reports\cite{trion-more-stbl1,trion-more-stbl2,stebe1,thilagam} showing trion states more stable than those of exciton regardless of the electron-hole mass ratio although the author of this paper have not obtained such results using a slightly different but more rigorous Hamiltonian. Not reaching a low energy after a few trials in variational methods does not rule out its actual existence and this problem will not be further probed here because it is off our main topic for this paper. Such issue aside, we have been familiar with the techniques needed for further development and let us now investigate the possible negative-energy state of tetron or 3E+1H. 

\section{The variational treatment for tetron}\label{sec-tetron}

\subsection{The Hamiltonian}

We assume that all particles have the same mass $m$.  The full Hamiltonian is
\begin{eqnarray}
H&=& - \frac{\hbar^2}{2m} ({\underline \nabla} _0^2 + {\underline \nabla} _1^2 +{\underline \nabla} _2^2 +{\underline \nabla} _3^2 ) \nonumber \\ 
&+& \frac{e^2}{4\pi \epsilon }(\frac{1}{{\underline r}_{12}}+\frac{1}{{\underline r}_{23}}+\frac{1}{{\underline r}_{31}}-\frac{1}{{\underline r}_{10}}-\frac{1}{{\underline r}_{20}}-\frac{1}{{\underline r}_{30}}),
\end{eqnarray}
where subscripts 1, 2, and 3 are indices to the three electrons while 0 designates the hole.  ${\underline r}_{12}$ is the distance between particles 1 and 2. The meanings of ${\underline r}_{23}$, ${\underline r}_{31}$,  ${\underline r}_{10}$,  ${\underline r}_{20}$, and  ${\underline r}_{30}$ are similarly defined. Using the underlined $r$ instead of pure $r$ is simply for notational purpose; we will in a moment transform the coordinates into a more convenient form. Under the current notation ${\underline \nabla} _0^2$ is given by 
\begin{equation}
{\underline \nabla} _0 ^2 = \frac{1}{{\underline r}_0}\frac{\partial}{\partial {\underline r}_0}({\underline r}_0\frac{\partial}{\partial {\underline r}_0}) +  \frac{1}{{\underline r}_0^2} \frac{\partial^2}{\partial{\underline \theta}_0 ^2}
\end{equation} 
and ${\underline \nabla} _1 ^2$, ${\underline \nabla} _2 ^2$, and  ${\underline \nabla} _3 ^2$ are understood similarly.
 
Performing the dimensionless unit transform same as made before in Sec. \ref{qm2d} , the Hamiltonian is now 

\begin{eqnarray}
H&=& - ({\underline \nabla} _0^2 + {\underline \nabla} _1^2 +{\underline \nabla} _2^2 +{\underline \nabla} _3^2 ) \nonumber \\ 
&&+ (\frac{1}{{\underline r}_{12}}+\frac{1}{{\underline r}_{23}}+\frac{1}{{\underline r}_{31}}-\frac{1}{{\underline r}_{10}}-\frac{1}{{\underline r}_{20}}-\frac{1}{{\underline r}_{30}}).
\end{eqnarray}

We introduce a new coordinate system 
\begin{eqnarray}
{\bf r}_0 &=& \frac{{\bf {\underline r}}_0+{\bf {\underline r}}_1+{\bf {\underline r}}_2+{\bf {\underline r}}_3}{4}, \nonumber \\
{\bf r}_1 &=& \frac{{\bf {\underline r}}_0+{\bf {\underline r}}_1-{\bf {\underline r}}_2-{\bf {\underline r}}_3}{2}, \nonumber \\
{\bf r}_2 &=& \frac{{\bf {\underline r}}_0-{\bf {\underline r}}_1+{\bf {\underline r}}_2-{\bf {\underline r}}_3}{2}, \nonumber \\
{\bf r}_3 &=& \frac{{\bf {\underline r}}_0-{\bf {\underline r}}_1-{\bf {\underline r}}_2+{\bf {\underline r}}_3}{2}, 
\end{eqnarray}
which are treated as vector transforms. 

These transforms result into the following relations,
\begin{eqnarray}
{\bf r}_0 - {\bf r}_1 &=& {\bf {\underline r}}_2+{\bf {\underline r}}_3, \nonumber \\
{\bf r}_0 - {\bf r}_2 &=& {\bf {\underline r}}_3+{\bf {\underline r}}_1, \nonumber \\
{\bf r}_0 - {\bf r}_3 &=& {\bf {\underline r}}_1+{\bf {\underline r}}_2, 
\end{eqnarray}
and 
\begin{eqnarray}
{\bf r}_1 - {\bf r}_2 &=& {\bf {\underline r}}_1-{\bf {\underline r}}_2, \nonumber \\
{\bf r}_2 - {\bf r}_3 &=& {\bf {\underline r}}_2-{\bf {\underline r}}_3, \nonumber \\
{\bf r}_3 - {\bf r}_1 &=& {\bf {\underline r}}_3-{\bf {\underline r}}_1. 
\end{eqnarray}

With the transformed coordinate system in place, the Hamiltonian now takes a new form,
 
\begin{eqnarray}
H&=& - (\frac{1}{4} \nabla _0^2 + \nabla _1^2 +\nabla _2^2 +\nabla _3^2 ) \nonumber \\ 
&+& (\frac{1}{r_{12}}+\frac{1}{r_{23}}+\frac{1}{r_{31}}-\frac{1}{r_{1+2}}-\frac{1}{r_{2+3}}-\frac{1}{r_{3+1}}). \label{4bh}
\end{eqnarray}
where for example
\begin{equation}
\nabla _0 ^2 = \frac{1}{r_0}\frac{\partial}{\partial r_0}(r_0\frac{\partial}{\partial r_0}) +  \frac{1}{r_0^2} \frac{\partial^2}{\partial \theta_0 ^2},
\end{equation} 
and 
\begin{equation}
r_{12}=\sqrt{r_1^2+r_2^2-2r_1r_2 \cos\theta _{12}},
\end{equation}
while $r_{1+2}$ is the short-hand for $|{\bf r}_1+{\bf r}_2|$ or $\sqrt{r_1^2+r_2^2+2r_1r_2 \cos\theta _{12}}$. 

${\bf r}_0$ now refers to the center-of-mass (c.m.). After discarding the irrelevant first term on RHS of Eq. (\ref{4bh}), we reduce the internal Hamiltonian in the c.m. frame to 
\begin{eqnarray}
H&=& - (\nabla _1^2 +\nabla _2^2 +\nabla _3^2 ) \nonumber \\ 
&&+ (\frac{1}{r_{12}}+\frac{1}{r_{23}}+\frac{1}{r_{31}}-\frac{1}{r_{1+2}}-\frac{1}{r_{2+3}}-\frac{1}{r_{3+1}}), \label{3bh}
\end{eqnarray} 
which is effectively a three-body problem.  

\subsection{Spin-singlet trial wavefunction of $1s$, $2p^+$, and $2p^-$ states}

It is easy to see that if the trial wavefunction for the three-body problem is antisymmetrized, the corresponding wavefunction for the our original four-body problem will be antisymmetric as well under the permutation of the original electron coordinates. Now we want to see how to make a three-body antisymmetrized wavefunctions.
   
The trial wavefunctions will be initially made only from the first four eigenstates for the purpose of simplicity. 
\begin{eqnarray}
 | 1s \rangle &=& \sqrt{\frac{2}{\pi}} \lambda \exp(-\lambda r), \nonumber  \\
  | 2s \rangle &=& \sqrt{\frac{2}{27\pi}}[\lambda -\frac{2}{3}\lambda ^2 r] \exp(-\frac{\lambda r}{3}),   \nonumber \\
  | 2p^{\pm} \rangle &=& \sqrt{\frac{4}{243\pi}} \lambda ^2 r \exp(-\frac{\lambda r}{3}) \exp(\pm i \theta).  \nonumber \\ \label{62} 
\end{eqnarray}
which are copied from the previous section for quick reference here. As a notation convenience which will be extensively used later, we also define  
\begin{equation}
 | 2p \rangle = \sqrt{\frac{4}{243\pi}} \lambda ^2 r \exp(-\frac{\lambda r}{3}). 
\end{equation}
Note that $\langle 2p | 2p \rangle = 1$ although $|2p\rangle$ is not necessarily orthogonal to other eigenstates. 

We now try to make an antisymmetrized wavefunction $\Psi$ out of Eqs. (\ref{62}).  $\Psi$ should satisfy our main object, $\langle \Psi|H |\Psi \rangle / \langle \Psi |\Psi \rangle  < 0$. 

The nice symmetry property of the Hamiltonian (\ref{3bh}) helps us rule out $\Psi $ only made of s states. To see this, we need to understand that the kinetic terms in Hamiltonian are positive definite, or $\langle \Psi|  - (\nabla _1^2 +\nabla _2^2 +\nabla _3^2 ) |\Psi \rangle / \langle \Psi |\Psi \rangle  > 0$. Thus the only possibility left for a negative energy state is by a negative Coulomb term $\langle \Psi|  (\frac{1}{r_{12}}+\frac{1}{r_{23}}+\frac{1}{r_{31}}-\frac{1}{r_{1+2}}-\frac{1}{r_{2+3}}-\frac{1}{r_{3+1}})  |\Psi \rangle / \langle \Psi |\Psi \rangle $. We only need to consider terms like  $\langle \Psi|  (\frac{1}{r_{12}}-\frac{1}{r_{1+2}})  |\Psi \rangle / \langle \Psi |\Psi \rangle $ because the total Coulomb energy is just three times as much because of the symmetry of $\Psi$ in ${\bf r}_1, {\bf r}_2, {\bf r}_3$. Now we show and exploit a further symmetry hidden in $r_{12}$ and $r_{1+2}$.

Note that 
\begin{eqnarray}
r_{12} &=& \sqrt{r_1^2+r_2^2-2r_1r_2 \cos\theta _{12}}, \\
r_{1+2} &=& \sqrt{r_1^2+r_2^2+2r_1r_2 \cos\theta _{12}} \\
 &=& \sqrt{r_1^2+r_2^2-2r_1r_2 \cos(\pi + \theta _{12})}.
\end{eqnarray}

Both $r_{12}$ and  $r_{1+2}$ are periodic in $\theta _1$ with a period $2\pi$. By the same argument we made in Sec. \ref{sec-trion} for two-body Coulomb integrals, we can integrate with respect to $\theta_{12}$ and $\theta_2$,  both from $0$ to $2\pi$. However $r_{1+2}$ is identical to $r_{12}$ except for a phase shift of $\pi$ in $\theta_{12}$. Since $\cos \theta_{12}$  has a period of $2\pi$ in $\theta_{12}$, $\langle \Psi|  \frac{1}{r_{12}}|\Psi \rangle / \langle \Psi |\Psi \rangle $ would be equal to $\langle \Psi| \frac{1}{r_{1+2}}  |\Psi \rangle / \langle \Psi |\Psi \rangle $ if $\Psi$ does not have any angular dependency such as those given by pure $s$ states. The total Coulomb energy would be 0 and we have no way whatsoever to obtain a negative energy. We will give a more general theorem later and exploit similar properties to greatly simplify our calculation. 

So one of the states chosen in $\Psi$ has to be $|2p^+\rangle$ or $|2p^-\rangle$. From our trial-and-error experience, we find that it is generally better to have two $2p$ states and one $s$ state.

As the first trial we want to test wavefunctions made out of $|1s\rangle$, $|2p^+\rangle$ and $|2p^-\rangle$.

By using variational approach, we also want to investigate the spin structure of such 3E+1H complex. Since there is no spin-orbit coupling term in the Hamiltonian, we need only to specify the total spin of three electrons for any trial wavefunction.   Basically there are two possible options for three electrons: either they are all lined up in the same direction or one of them is against the other two. The hole is also a fermion, whose spin tends to be opposite to at least one of the electrons. Thus the total spin of this complex is either 0 or $\hbar$, corresponding to spin singlet or triplet states. We would consider this explicitly in our calculation. 

First let us start with three orthogonal states $| 1s\updownarrow \rangle$,  $| 2p^+\uparrow \rangle$ and $| 2p^-\downarrow \rangle$. The arrows designate spin states. The complex is now in a spin singlet state because the mixed spin wavefunction $| \updownarrow \rangle = \frac{1}{\sqrt{2}}| \uparrow \rangle + \frac{1}{\sqrt{2}}| \downarrow \rangle $ should combine with that of the hole to give a singlet for one electron-hole pair. Writing  $| \updownarrow \rangle$ also means that we do not care about the exact spin state; it can be either up or down. 

The corresponding Slater determinant is 
\begin{eqnarray}
&&\sqrt{\frac{1}{6}}\left| 
\begin{array}{ccc}
| 1s_1\updownarrow_1 \rangle &| 2p_1^+\uparrow_1 \rangle &| 2p_1^-\downarrow _1 \rangle \\
| 1s_2\updownarrow_2 \rangle &| 2p_2^+\uparrow_2 \rangle &| 2p_2^-\downarrow _2 \rangle \\
| 1s_3\updownarrow_3 \rangle &| 2p_3^+\uparrow_3 \rangle &| 2p_3^-\downarrow _3 \rangle \\
\end{array}
\right| \nonumber \\
&&=  
\sqrt{\frac{1}{6}}\left| 
\begin{array}{cc}
  | 2p_2^+\uparrow_2 \rangle & | 2p_2^-\downarrow_2 \rangle \\
  | 2p_3^+\uparrow_3 \rangle & | 2p_3^-\downarrow_3 \rangle \\
\end{array}
\right| | 1s_1\updownarrow_1 \rangle
+
\sqrt{\frac{1}{6}}\left| 
\begin{array}{cc}
  | 2p_1^+\uparrow_1 \rangle & | 2p_1^-\downarrow_1 \rangle \\
  | 2p_2^+\uparrow_2 \rangle & | 2p_2^-\downarrow_2 \rangle \\
\end{array}
\right| | 1s_3\updownarrow_3 \rangle
-
\sqrt{\frac{1}{6}}\left| 
\begin{array}{cc}
  | 2p_1^+\uparrow_1 \rangle & | 2p_1^-\downarrow_1 \rangle \\
  | 2p_3^+\uparrow_3 \rangle & | 2p_3^-\downarrow_3 \rangle \\
\end{array}
\right| | 1s_2\updownarrow_2 \rangle  \nonumber \\
&&{}= \sqrt{\frac{1}{6}} \bbox\{
 | 2p_22p_3 \rangle \{ \exp [i(\theta_2 - \theta_3)]|\uparrow_2 \downarrow_3 \rangle - \exp [i(\theta_3 - \theta_2)]|\uparrow_3 \downarrow_2 \rangle \} | 1s_1\updownarrow_1 \rangle 
\nonumber \\
&&\phantom{{}= \sqrt{\frac{1}{6}} \bbox\{}+| 2p_12p_2 \rangle \{ \exp [i(\theta_1 - \theta_2)]|\uparrow_1 \downarrow_2 \rangle - \exp [i(\theta_2 - \theta_1)]|\uparrow_2 \downarrow_1 \rangle \} | 1s_3\updownarrow_3 \rangle
\nonumber \\
&&\phantom{{}= \sqrt{\frac{1}{6}} \bbox\{}-| 2p_12p_3 \rangle \{ \exp [i(\theta_1 - \theta_3)]|\uparrow_1 \downarrow_3 \rangle - \exp [i(\theta_3 - \theta_1)]|\uparrow_3 \downarrow_1 \rangle \} | 1s_2\updownarrow_2 \rangle
\bbox\}, \label{slater1}
\end{eqnarray}
where for example $\updownarrow _1$ indicates a mixed spin state for electron 1 while $2p_2^+$ says that electron 2 is in a $2p^+$ state and so on so forth.  Subscripts 1, 2, and 3 as usual are electron indices.

There will be 6 terms in this antisymmetrized trial wavefunction, which would render follow-on manipulation difficult. Here we do a small trick by adding to the above the following Slater determinant, which is similar to Eq. (\ref{slater1}) except for the switch between $2p^+$ and $2p^-$ states. 
\begin{eqnarray}
&&\sqrt{\frac{1}{6}}\left| 
\begin{array}{ccc}
| 1s_1\updownarrow_1 \rangle &| 2p_1^-\uparrow_1 \rangle &| 2p_1^+\downarrow _1 \rangle \\
| 1s_2\updownarrow_2 \rangle &| 2p_2^-\uparrow_2 \rangle &| 2p_2^+\downarrow _2 \rangle \\
| 1s_3\updownarrow_3 \rangle &| 2p_3^-\uparrow_3 \rangle &| 2p_3^+\downarrow _3 \rangle \\
\end{array}
\right| \nonumber \\
&&=  
\sqrt{\frac{1}{6}}\left| 
\begin{array}{cc}
  | 2p_2^-\uparrow_2 \rangle & | 2p_2^+\downarrow_2 \rangle \\
  | 2p_3^-\uparrow_3 \rangle & | 2p_3^+\downarrow_3 \rangle \\
\end{array}
\right| | 1s_1\updownarrow_1 \rangle
+
\sqrt{\frac{1}{6}}\left| 
\begin{array}{cc}
  | 2p_1^-\uparrow_1 \rangle & | 2p_1^+\downarrow_1 \rangle \\
  | 2p_2^-\uparrow_2 \rangle & | 2p_2^+\downarrow_2 \rangle \\
\end{array}
\right| | 1s_3\updownarrow_3 \rangle
-
\sqrt{\frac{1}{6}}\left| 
\begin{array}{cc}
  | 2p_1^-\uparrow_1 \rangle & | 2p_1^+\downarrow_1 \rangle \\
  | 2p_3^-\uparrow_3 \rangle & | 2p_3^+\downarrow_3 \rangle \\
\end{array}
\right| | 1s_2\updownarrow_2 \rangle  \nonumber \\
&&{}= \sqrt{\frac{1}{6}} \bbox\{
 | 2p_22p_3 \rangle \{ \exp [i(\theta_3 - \theta_2)]|\uparrow_2 \downarrow_3 \rangle - \exp [i(\theta_2 - \theta_3)]|\uparrow_3 \downarrow_2 \rangle \} | 1s_1\updownarrow_1 \rangle 
\nonumber \\
&&\phantom{{}= \sqrt{\frac{1}{6}} \bbox\{}+| 2p_12p_2 \rangle \{ \exp [i(\theta_2 - \theta_1)]|\uparrow_1 \downarrow_2 \rangle - \exp [i(\theta_1 - \theta_2)]|\uparrow_2 \downarrow_1 \rangle \} | 1s_3\updownarrow_3 \rangle
\nonumber \\
&&\phantom{{}= \sqrt{\frac{1}{6}} \bbox\{}-| 2p_12p_3 \rangle \{ \exp [i(\theta_3 - \theta_1)]|\uparrow_1 \downarrow_3 \rangle - \exp [i(\theta_1 - \theta_3)]|\uparrow_3 \downarrow_1 \rangle \} | 1s_2\updownarrow_2 \rangle
\bbox\}. \label{slater2}
\end{eqnarray}

The result is the following after normalization, 
\begin{eqnarray}
\Psi =  \frac{2}{\sqrt{6}} \{ && |2p_1 2p_2 1s_3\rangle \cos \theta_{12} |{\cal S}^a_{12,3}\rangle \nonumber \\
&&+ |2p_2 2p_3 1s_1\rangle \cos \theta_{23} |{\cal S}^a_{23,1}\rangle \nonumber \\
&&+ |2p_3 2p_1 1s_2\rangle \cos \theta_{31} |{\cal S}^a_{31,2}\rangle
\},
\end{eqnarray}
where the spin wavefunction $|{\cal S}^a_{12,3}\rangle = \frac{\sqrt{2}}{2}(|\uparrow_1 \downarrow_2 \rangle  -| \uparrow_2 \downarrow_1 \rangle)|\updownarrow_3\rangle $ which is antisymmetric in the first two indices 1 and 2, and note that 
\begin{equation}
\langle {\cal S}^a_{12,3}| {\cal S}^a_{23,1} \rangle = \langle {\cal S}^a_{12,3}| {\cal S}^a_{31,2} \rangle = 
  \langle {\cal S}^a_{23,1}| {\cal S}^a_{31,2} \rangle =  -\frac{1}{2}.
\end{equation}

$\Psi$ is already normalized and thus $\langle \Psi | \Psi \rangle = 1$. The kinetic energy of this trial wavefunction $\langle \Psi |- \nabla _1^2 - \nabla _2^2  - \nabla _3^2 | \Psi \rangle $ is easy to work out 
\begin{equation}
\langle \Psi |- \nabla _1^2 - \nabla _2^2  - \nabla _3^2 | \Psi \rangle = \frac{11}{9} \lambda ^2.
\end{equation}
Actually this can be figured out without explicit calculation. We have three particles in $1s$, $2p^+$ and $2p^-$ states whose kinetic energy are respectively $\lambda ^2$, $\lambda ^2 /9 $  and $\lambda ^2 /9 $, thus the total kinetic energy should be the sum, which gives   $11\lambda ^2/9$. 

As for the Coulomb energy, we need only to calculate this term
\begin{equation}
\langle \Psi |\frac{1}{r_{12}}-\frac{1}{r_{1+2}} | \Psi \rangle,
\end{equation}
because 
\begin{equation}
\langle \Psi |\frac{1}{r_{12}}-\frac{1}{r_{1+2}} | \Psi \rangle = 
\langle \Psi |\frac{1}{r_{23}}-\frac{1}{r_{2+3}} | \Psi \rangle = 
\langle \Psi |\frac{1}{r_{31}}-\frac{1}{r_{3+1}} | \Psi \rangle,
\end{equation}
considering symmetry in electron permutation. 

We then write explicitly 
\begin{eqnarray}
&&\langle \Psi |\frac{1}{r_{12}}-\frac{1}{r_{1+2}} | \Psi \rangle \nonumber \\ 
&&{}=\frac{4}{6} \{ \langle 2p_12p_21s_3 |\cos ^2 \theta_{12} (\frac{1}{r_{12}}-\frac{1}{r_{1+2}}) | 2p_12p_21s_3 \rangle  \nonumber \\ 
&&\phantom{{}=\frac{4}{6} \{} + \langle 2p_22p_31s_1 |\cos ^2 \theta_{23} (\frac{1}{r_{12}}-\frac{1}{r_{1+2}}) | 2p_22p_31s_1 \rangle  \nonumber \\
&&\phantom{{}=\frac{4}{6} \{} + \langle 2p_32p_11s_2 |\cos ^2 \theta_{31} (\frac{1}{r_{12}}-\frac{1}{r_{1+2}}) | 2p_32p_11s_2 \rangle  \nonumber \\
&&\phantom{{}=\frac{4}{6} \{} - \langle 2p_22p_31s_1 |\cos \theta_{12} \cos \theta_{23}  (\frac{1}{r_{12}}-\frac{1}{r_{1+2}}) | 2p_12p_21s_3 \rangle  \nonumber \\ 
&&\phantom{{}=\frac{4}{6} \{} - \langle 2p_32p_11s_2 |\cos \theta_{12} \cos \theta_{31}  (\frac{1}{r_{12}}-\frac{1}{r_{1+2}}) | 2p_12p_21s_3 \rangle  \nonumber \\
&&\phantom{{}=\frac{4}{6} \{} - \langle 2p_32p_11s_2 |\cos \theta_{23} \cos \theta_{31}  (\frac{1}{r_{12}}-\frac{1}{r_{1+2}}) | 2p_32p_21s_1 \rangle \}.  \label{r12r1+2-1}
\end{eqnarray}

For obvious identities from symmetry
\begin{eqnarray}
&& \langle 2p_22p_31s_1 |\cos ^2 \theta_{23} (\frac{1}{r_{12}}-\frac{1}{r_{1+2}}) | 2p_22p_31s_1 \rangle  \nonumber \\
&& = \langle 2p_32p_11s_2 |\cos ^2 \theta_{31} (\frac{1}{r_{12}}-\frac{1}{r_{1+2}}) | 2p_32p_11s_2 \rangle,  \nonumber \\
&& \langle 2p_22p_31s_1 |\cos \theta_{12} \cos \theta_{23}  (\frac{1}{r_{12}}-\frac{1}{r_{1+2}}) | 2p_12p_21s_3 \rangle  \nonumber \\ 
&& = \langle 2p_32p_11s_2 |\cos \theta_{12} \cos \theta_{31}  (\frac{1}{r_{12}}-\frac{1}{r_{1+2}}) | 2p_12p_21s_3 \rangle,   \nonumber 
\end{eqnarray}
Eq. (\ref{r12r1+2-1}) is simplified to 
\begin{eqnarray}
&&\langle \Psi |\frac{1}{r_{12}}-\frac{1}{r_{1+2}} | \Psi \rangle \nonumber \\
&&{}=\frac{4}{6} \{ \langle 2p_12p_21s_3 |\cos ^2 \theta_{12} (\frac{1}{r_{12}}-\frac{1}{r_{1+2}}) | 2p_12p_21s_3 \rangle  \nonumber \\ 
&&\phantom{{}=\frac{4}{6} \{} +2 \langle 2p_22p_31s_1 |\cos ^2 \theta_{23} (\frac{1}{r_{12}}-\frac{1}{r_{1+2}}) | 2p_22p_31s_1 \rangle  \nonumber \\
&&\phantom{{}=\frac{4}{6} \{}-2 \langle 2p_22p_31s_1 |\cos \theta_{12} \cos \theta_{23}  (\frac{1}{r_{12}}-\frac{1}{r_{1+2}}) | 2p_12p_21s_3 \rangle  \nonumber \\ 
&&\phantom{{}=\frac{4}{6} \{} - \langle 2p_32p_11s_2 |\cos \theta_{23} \cos \theta_{31}  (\frac{1}{r_{12}}-\frac{1}{r_{1+2}}) | 2p_22p_31s_1 \rangle \}.  \label{r12r1+2-2}
\end{eqnarray}

The first term on RHS of Eq. (\ref{r12r1+2-2}) will be zero. To understand this, we prove the following theorem. 

{\it Theorem}: Suppose that $f_1(r_1,r_2,r_3)$ and  $f_2(r_1,r_2,r_3)$ are functions only of $r_1$, $r_2$, and $r_3$ ({\it i.e.}, do not depend on $\theta_1$, $\theta_2$, and $\theta_3$). Also assume that $\langle f_1|\frac{1}{r_{12}}|f_2 \rangle$ exists where $f_1$ and $f_2$ are short-hands for  $f_1(r_1,r_2,r_3)$ and  $f_2(r_1,r_2,r_3)$. Then we have 

\begin{eqnarray}
&&\langle f_1 |\cos^n \theta_{12}(\frac{1}{r_{12}}-\frac{1}{r_{1+2}})|f_2 \rangle \nonumber \\
&&=\cases{
   & $0 $,  if $n$ is even; \cr 
   & $\langle f_1 |2\cos^n \theta_{12}(\frac{1}{r_{12}})|f_2 \rangle$,  if n is odd. \cr} 
\end{eqnarray}

{\it Proof}: 
\begin{eqnarray}
&&\langle f_1 |\cos^n \theta_{12}(\frac{1}{r_{12}}-\frac{1}{r_{1+2}})|f_2 \rangle \nonumber \\
&&= \int _0^ {\infty} r_1 dr_1 \int _0^ {\infty} r_2 dr_2 \int _0^ {\infty} r_3 dr_3 \int _0^ {2\pi} d \theta _1 \int _0^ {2\pi} d \theta _2 \int _0^ {2\pi} d \theta _3 f_1  f_2  \cos^n \theta_{12}(\frac{1}{r_{12}}-\frac{1}{r_{1+2}}) \nonumber \\ 
&&= \int _0^ {\infty} r_1 dr_1 \int _0^ {\infty} r_2 dr_2 \int _0^ {\infty} r_3 dr_3 \int _0^ {2\pi} d \theta _3 \int _0^ {2\pi} d \theta _2 \int _0^ {2\pi} d \theta _{12} f_1  f_2  \cos^n \theta_{12}(\frac{1}{r_{12}}-\frac{1}{r_{1+2}}) \label{77}
\end{eqnarray}
where in the second step we have changed variables from $\theta_1$ to $\theta_{12}$ for the same reason that we discussed before in Sec. \ref{sec-trion}. We consider first the integration with respect to  $\theta_{12}$ in Eq. (\ref{77}),  or the following expression
\begin{eqnarray}
&&\int _0^ {2\pi} d \theta _{12} \cos^n \theta_{12}(\frac{1}{r_{12}}-\frac{1}{r_{1+2}}) \nonumber \\
&&= \int _0^ {2\pi} d \theta _{12} \cos^n \theta_{12}(\frac{1}{\sqrt{r_1^2+r_2^2-2r_1r_2\cos\theta_{12}}} - \frac{1}{\sqrt{r_1^2+r_2^2-2r_1r_2\cos(\pi + \theta_{12})}}). \label{r12r1+2theta12}
\end{eqnarray}
The last term on RHS of Eq. (\ref{r12r1+2theta12}) is 
\begin{equation}
-\int _0^ {2\pi} d \theta _{12} \cos^n \theta_{12} \frac{1}{\sqrt{r_1^2+r_2^2-2r_1r_2\cos(\pi + \theta_{12})}}.
\end{equation}
Changing variable from $\theta_{12}$ to an arbitrary $x= \pi + \theta_{12} $ in this integral, we have
\begin{eqnarray}
&&-\int _0^ {2\pi} d \theta _{12} \cos^n \theta_{12} \frac{1}{\sqrt{r_1^2+r_2^2-2r_1r_2\cos(\pi + \theta_{12})}}  \nonumber \\ 
&&= - \int _{\pi}^ {3\pi} d x  \cos^n (x-\pi) \frac{1}{\sqrt{r_1^2+r_2^2-2r_1r_2\cos x}} \nonumber \\
&&= - \int _{\pi}^ {3\pi} d x  (-1)^n \cos^n x \frac{1}{\sqrt{r_1^2+r_2^2-2r_1r_2\cos x}}. \label{r12r1+2x}
\end{eqnarray}
Because the integrand on RHS of Eq. (\ref{r12r1+2x}) is a function of $x$ with a period $2\pi$, integrating from $0$ to $\pi$ is equal to integrating from $2\pi$ to $3\pi$, or
\begin{eqnarray}
&& -\int _{2\pi}^ {3\pi} d x (-1)^n \cos^n x \frac{1}{\sqrt{r_1^2+r_2^2-2r_1r_2\cos x}}  \nonumber \\
&&= - \int _{0}^ {\pi} d x (-1)^n \cos^n x \frac{1}{\sqrt{r_1^2+r_2^2-2r_1r_2\cos x}},
\end{eqnarray} 
or alternatively  
\begin{eqnarray}
&& -\int _{\pi}^ {3\pi} d x (-1)^n \cos^n x \frac{1}{\sqrt{r_1^2+r_2^2-2r_1r_2\cos x}} \nonumber \\
&&= -\int _{0}^ {2\pi} d x  (-1)^n \cos^n x \frac{1}{\sqrt{r_1^2+r_2^2-2r_1r_2\cos x}}. \label{r12r1+2x-0-3pi}
\end{eqnarray}

After Eq. (\ref{r12r1+2x-0-3pi}) is substituted in Eq. (\ref{r12r1+2x}), 
\begin{eqnarray}
&&-\int _0^ {2\pi} d \theta _{12} \cos^n \theta_{12} \frac{1}{\sqrt{r_1^2+r_2^2-2r_1r_2\cos(\pi + \theta_{12})}}  \nonumber \\ 
&&= -\int _{0}^ {2\pi} d x  (-1)^n \cos^n x \frac{1}{\sqrt{r_1^2+r_2^2-2r_1r_2\cos x }}  \nonumber \\ 
&&= -\int _{0}^ {2\pi} d \theta _{12}  (-1)^n \cos^n  \theta _{12} \frac{1}{\sqrt{r_1^2+r_2^2-2r_1r_2\cos\theta _{12}}}, \label{r12r1+2x-0-3pi-2}
\end{eqnarray}
where in the second step variable $x$ is relabelled as $\theta _{12}$.

Inserting Eq. (\ref{r12r1+2x-0-3pi-2}) into (\ref{r12r1+2theta12}), we have
\begin{eqnarray}
&&\int _0^ {2\pi} d \theta _{12} \cos^n \theta_{12}(\frac{1}{r_{12}}-\frac{1}{r_{1+2}}) \nonumber \\
&&= \int _0^ {2\pi} d \theta _{12} \cos^n \theta_{12} [\frac{1}{\sqrt{r_1^2+r_2^2-2r_1r_2\cos\theta_{12}}}- \frac{(-1)^n }{\sqrt{r_1^2+r_2^2-2r_1r_2\cos\theta_{12}}}]  \nonumber \\
&&= \cases{
   & $0 $,  if $n$ is even; \cr                            
   & $\int _0^ {2\pi} d \theta _{12} \frac{2\cos^n \theta_{12}}{r_{12}}$,  if $n$ is odd. \cr} \label{84}
\end{eqnarray}

Then we substitute Eq. (\ref{84}) back to Eq. (\ref{77}) to give the theorem. This concludes our proof.

The second term on RHS of Eq. (\ref{r12r1+2-2}) or  $2 \langle 2p_22p_31s_1 |\cos \theta^2_{23} (\frac{1}{r_{12}}-\frac{1}{r_{1+2}})  | 2p_22p_31s_1 \rangle$ has the factor 
\begin{eqnarray}
\cos^2\theta_{23} &=& [\cos \theta_2 \cos \theta_3 +\sin \theta_2 \sin \theta_3 ]^2 \nonumber \\
&=& \cos^2 \theta_2 \cos^2 \theta_3  + \sin^2 \theta_2 \sin^2 \theta_3  \nonumber \\
&&+ 2 \cos \theta_2 \cos \theta_3 \sin \theta_2 \sin \theta_3,
\end{eqnarray}
where last term on the right gives 0 after integration with respect to $\theta_3$, while the first two terms, combined with $\frac{1}{r_{12}} - \frac{1}{r_{1+2}}$ are 0 after integration with respect to $\theta_{12}$ from our theorem, because there is no odd power of $\cos \theta_{12}$ in the integrand. 

The third term on RHS of Eq. (\ref{r12r1+2-2}) is 0 after integration with respect to $\theta_3$ and we have only the last term left. 

Because 
\begin{eqnarray}
&\cos & \theta_{23} \cos \theta_{31}   \nonumber \\
&=& [\cos \theta_2 \cos \theta_3 +\sin \theta_2 \sin \theta_3 ] [\cos \theta_3 \cos \theta_1 +\sin \theta_3 \sin \theta_1 ] \nonumber \\
&=& \cos \theta_2 \cos \theta_1  \cos^2 \theta_3  + \sin \theta_2 \sin \theta_1 \sin^2 \theta_3 \nonumber \\
&&+ \cos \theta_3 \cos \theta_1 \sin \theta_2 \sin \theta_3  + \cos \theta_2 \cos \theta_3 \sin \theta_3 \sin \theta_1,  
\end{eqnarray}
which after integration with respect to $\theta_3$ gives 
\begin{eqnarray}
&&\pi (\cos \theta_2 \cos \theta_1  +  \sin \theta_2 \sin \theta_1)  \nonumber \\
&&= \pi\cos \theta_{12}.
\end{eqnarray}

Thus  
\begin{eqnarray}
&& - \langle 2p_32p_11s_2 |\cos \theta_{23} \cos \theta_{31}  (\frac{1}{r_{12}}-\frac{1}{r_{1+2}}) | 2p_32p_21s_1 \rangle \nonumber \\
&& = - \langle 2p_11s_2| \frac{\cos \theta_{12}}{r_{12}} | 2p_21s_1 \rangle
\end{eqnarray}
which is worked out in the same fashion as integral (\ref{int12}) in section \ref{sec-trion} to give
\begin{equation}
-\langle 2p_11s_2|\frac{\cos \theta_{12}}{r_{12}} | 2p_21s_1 \rangle = -0.0172573 \lambda .
\end{equation}

The total Coulomb energy is thus $-0.0345146 \lambda$ and we have 
\begin{eqnarray}
E &=&\frac {\langle \Psi |H| \Psi \rangle}{\langle \Psi|\Psi\rangle} \nonumber \\
&=& \frac{11\lambda ^2}{9} - 0.0345146 \lambda  \nonumber \\
&=& \frac{11}{9} (\lambda - 0.0141196)^2 - 0.00019936,
\end{eqnarray}
from which the upper bound for ground-state energy is therefore $-0.00019936$. 

This confirms our conjecture that 3E+1H indeed forms a quasi-bound state. Here quasi-bound merely means that the 4 particles are relatively localized or near one to another and the total energy of this system is {\em lower} than when they are infinitely apart. Note that the 4 particle as a whole could still move freely with an effective mass of $4m$.  

We proceed on to try other wavefunctions and want to improve our result. 

\subsection{Spin-triplet trial wavefunction of $1s$, $2p^+$, and $2p^-$ states}

The next logic choice is still $1s$, $2p^+$, and $2p^-$ states but with a spin-triplet configuration, 
\begin{eqnarray}
\Psi = \frac{2}{\sqrt{6}} \{&&|2p_1 2p_2 1s_3\rangle \sin \theta_{12} |{\cal S}^s_{1,23}\rangle \nonumber \\
&&+ |2p_2 2p_3 1s_1\rangle \sin \theta_{23} |{\cal S}^s_{23,1}\rangle \nonumber \\
&&+ |2p_3 2p_1 1s_2\rangle \sin \theta_{31} |{\cal S}^s_{31,2}\rangle
\},
\end{eqnarray}
where the normalized spin wavefunction $|{\cal S}^s_{12,3}\rangle$ is symmetrized in indices 1 and 2, and describes a spin triplet state for electrons 1 and 2.  $|{\cal S}^s_{123}\rangle$ may be written $ |\uparrow_1 \uparrow_2 \updownarrow_3 \rangle = \frac{\sqrt{2}}{2}|\uparrow_1 \uparrow_2 \rangle  (|\uparrow_3\rangle + |\downarrow_3\rangle)$ but can also take other forms, for example $|\downarrow_1 \downarrow_2 \updownarrow_3 \rangle $ or $|\frac{1}{\sqrt{2}}(\uparrow_1 \downarrow_2 + \uparrow_2 \downarrow_1 )\updownarrow_3 \rangle$. Three forms of $|{\cal S}^s_{12,3}\rangle$ all satisfy $\langle {\cal S}^s_{12,3} | {\cal S}^s_{12,3} \rangle$ =1 and $\langle {\cal S}^s_{12,3} | {\cal S}^s_{23,1} \rangle = 1/2$ and so on. 

This trial wavefunction gives exactly the same result as the spin singlet trial wavefunction, 
\begin{eqnarray}
E &=&\frac {\langle \Psi |H| \Psi \rangle}{\langle \Psi|\Psi\rangle} \nonumber \\
&=& \frac{11\lambda ^2}{9} - 0.0345146 \lambda \nonumber \\
&=& \frac{11}{9} (\lambda - 0.0141196)^2 - 0.00019936,
\end{eqnarray} 
which gives $E_{min}= -0.00019936$. We are not yet able to tell whether the spin singlet or the triplet state is more stable. 

\subsection{Spin-singlet trial wavefunction of modified $1s$, $2p^+$, and $2p^-$ states}

We know in such a 3E+1H complex, the strong repulsive Coulomb force is a huge penalty we have pay attention to, thus it would be nice if our trial wavefunction is small when two electron coordinates are close. Take the spin-singlet wavefunction for example,

\begin{eqnarray}
\Psi = \frac{2}{\sqrt{6}} \{&&|2p_1 2p_2 1s_3\rangle \cos \theta_{12} |{\cal S}^a_{12,3}\rangle \nonumber \\
&&+ |2p_2 2p_3 1s_1\rangle \cos \theta_{23} |{\cal S}^a_{23,1}\rangle \nonumber \\
&&+ |2p_3 2p_1 1s_2\rangle \cos \theta_{31} |{\cal S}^a_{31,2}\rangle
\}, \label{91}
\end{eqnarray}
if we can modify $\Psi$ so that it is small when two electrons are close, the total energy then might be lower. So we choose the following, 

\begin{eqnarray}
\Psi = \frac{2}{\sqrt{6}} \{&& |2p_1 2p_2 1s_3\rangle (1-\cos \theta_{12}) |{\cal S}^a_{12,3}\rangle \nonumber \\
&&+ |2p_2 2p_3 1s_1\rangle (1- \cos \theta_{23}) |{\cal S}^a_{23,1}\rangle \nonumber \\
&&+ |2p_3 2p_1 1s_2\rangle (1- \cos \theta_{31}) |{\cal S}^a_{31,2}\rangle
\}, \label{94}
\end{eqnarray}
where the first term on RHS will be zero when $\theta _{12} = 0$, or $\theta_1 = \theta_2 $, or in other words, when electrons 1 and 2 are near. 

Note that after this modification, the three particles are no longer in pure $1s$, $2p^+$, and $2p^-$ states, but we may still use these words for notational purpose, and call them modified $1s$, $2p^+$, and $2p^-$ states.

Then 
\begin{eqnarray}
\langle \Psi | \Psi \rangle =&& 2 \langle 2p_1 2p_2 1s_3 |1- 2 \cos \theta_{12} + \cos^2 \theta_{12} |  2p_1 2p_2 1s_3 \rangle \nonumber \\
&&-  2 \langle 2p_1 2p_2 1s_3 |1- \cos \theta_{12} - \cos \theta_{23} + \cos \theta_{12} \cos \theta_{23} |  2p_2 2p_3 1s_1 \rangle \nonumber \\
=&& 3 - 2\{ \langle 2p_1 1s_3 | 2p_3 1s_1 \rangle + \langle 2p_1 2p_2 1s_3 | \cos \theta_{12} \cos \theta_{23} |  2p_2 2p_3 1s_1 \rangle   \},
\end{eqnarray}
where we have used identities
\begin{eqnarray}
&&\langle 2p_1 2p_2 1s_3 |1- 2 \cos \theta_{12} + \cos^2 \theta_{12} |  2p_1 2p_2 1s_3 \rangle \nonumber \\
&&= \langle 2p_2 2p_3 1s_1 |1- 2 \cos \theta_{23} + \cos^2 \theta_{23} |  2p_2 2p_3 1s_1 \rangle \nonumber \\
&&= \langle 2p_3 2p_1 1s_2 |1- 2 \cos \theta_{31} + \cos^2 \theta_{31} |  2p_3 2p_1 1s_2 \rangle, \label{96}
\end{eqnarray}
and 
\begin{eqnarray}
&&\langle 2p_1 2p_2 1s_3 |1- \cos \theta_{12} - \cos \theta_{23} + \cos \theta_{12} \cos \theta_{23} |  2p_2 2p_3 1s_1 \rangle \nonumber \\
&&= \langle 2p_2 2p_3 1s_1 |1- \cos \theta_{23} - \cos \theta_{31} + \cos \theta_{23} \cos \theta_{31} |  2p_3 2p_1 1s_2 \rangle, \label{97}
\end{eqnarray}
and the like. 

The exact value of $\langle \Psi | \Psi \rangle$ is found to be $90/32$. 

Next we want to find out the kinetic energy 
\begin{eqnarray}
&&\langle \Psi | - \nabla _1^2 - \nabla _2^2 - \nabla _3^2   |\Psi \rangle \nonumber \\
&&= 3 \langle \Psi | - \nabla _1^2 |\Psi \rangle \nonumber \\
&&= 3 \langle   \nabla _1 \Psi  | \nabla _1 \Psi \rangle,
\end{eqnarray}
where 
\begin{equation}
\nabla _1 = {\bf {\hat r}}_1 \frac{\partial}{\partial r_1} + {\bbox{\hat \theta}}_1 \frac{1}{r_1} \frac{\partial}{\partial \theta _1}.
\end{equation} 

Because
\begin{eqnarray}
\nabla _1  |\Psi \rangle = \frac{2}{\sqrt{6}} \{&(&\frac{1}{r_1} -\frac{\lambda}{3}) | 2p_1 2p_2 1s_3 \rangle (1-\cos \theta_{12}) {\cal S}^a_{12,3} \nonumber \\
&& -\lambda | 2p_2 2p_3 1s_1 \rangle (1-\cos \theta_{23}) {\cal S}^a_{23,1} \nonumber \\
&& + (\frac{1}{r_1} -\frac{\lambda}{3}) | 2p_3 2p_1 1s_2 \rangle (1-\cos \theta_{31}) {\cal S}^a_{31,2} \} {\bf {\hat r}}_1  \nonumber \\
&&{} + \frac{2}{\sqrt{6}}\frac{{\bbox{\hat \theta}}_1}{r_1} \{ | 2p_1 2p_2 1s_3 \rangle \sin \theta _{12}  {\cal S}^a_{12,3} \nonumber \\
&& \phantom{{} + \frac{2}{\sqrt{6}}\frac{{\bbox{\hat \theta}}_1}{r_1} \{} - | 2p_3 2p_1 1s_2 \rangle \sin \theta _{31}  {\cal S}^a_{31,2} \},
\end{eqnarray}
we have
\begin{eqnarray}
&&\langle   \nabla _1 \Psi  | \nabla _1 \Psi \rangle \nonumber \\
&&{}= \frac{4}{6} \{ 2 \langle 2p_1 2p_2 | (\frac{1}{r_1^2} - \frac{2\lambda}{3r_1} + \frac{\lambda^2}{9})(1-2\cos\theta_{12}+\cos^2\theta_{12}) | 2p_1 2p_2  \rangle  \nonumber \\
&& \phantom{{}= \frac{4}{6} \{} + \langle 2p_2 2p_3 |\lambda^2(1+\cos^2\theta_{23}-2\cos\theta_{23}) | 2p_2 2p_3  \rangle  \nonumber \\
&& \phantom{{}= \frac{4}{6} \{} -2  \langle 2p_1 1s_3 | (\frac{\lambda^2}{3}-\frac{\lambda}{r_1}) | 1s_1 2p_3  \rangle  \nonumber \\
&&  \phantom{{}= \frac{4}{6} \{}-  \langle 2p_1 2p_2 1s_3| (\frac{1}{r_1^2} - \frac{2\lambda}{3r_1} + \frac{\lambda^2}{9})| 2p_1 2p_3 1s_2 \rangle  \nonumber \\
&&  \phantom{{}= \frac{4}{6} \{}+  2 \langle 2p_1 2p_2 | \frac{\sin^2\theta_{12}}{r_1^2} | 2p_1 2p_2  \rangle  \},
\end{eqnarray}
where we have used symmetry identities similar to Eqs. (\ref{96}) and (\ref{97})to merge some terms. Those terms which go to 0 after integration of one variable, say $\theta_2$,  are dropped as well. 

After somewhat tedious but straightforward calculation, the total kinetic energy is $\frac{1405}{432} \lambda^2$ 

The Coulomb energy of the trial wavefunction is calculated as usual though with added complexity. We just give the result here, $-0.566523 \lambda$. 

Thus  
\begin{eqnarray}
E &=&\frac {\langle \Psi |H| \Psi \rangle}{\langle \Psi|\Psi\rangle} \nonumber \\
&=& 1.15638 \lambda ^2  - 0.201430 \lambda  \nonumber \\
&=&   1.15638 (\lambda - 0.0870951)^2 - 0.00877178,
\end{eqnarray}
which gives $E_{min} = - 0.00877178$, a much improved result over previous cases. 

Also recall that we have spin-singlet state in this case. In contrast, the spin-triplet trial wavefunction of Eq. (\ref{91}) cannot be so modified as Eq. (\ref{94}) because the first term on RHS of Eq. (\ref{91}) is already small when electrons 1 and 2 are close. This suggests that antiparallel spins be energetically more favorable than parallel ones. We will from now on focus our attention to spin-singlet state only. 

\subsection{Spin-singlet trial wavefunction of modified $2s$, $2p^+$, and $2p^-$ states}

The next trial wavefunction and the final one in this paper is made of modified $2s$, $2p^+$, and $2p^-$ states, 
\begin{eqnarray}
\Psi = \frac{2}{\sqrt{6}} \{ &&|2p_1 2p_2 2s_3\rangle (1-\cos \theta_{12}) |{\cal S}^a_{12,3}\rangle \nonumber \\
&&+ |2p_2 2p_3 2s_1\rangle (1- \cos \theta_{23}) |{\cal S}^a_{23,1}\rangle \nonumber \\
&&+ |2p_3 2p_1 2s_2\rangle (1- \cos \theta_{31}) |{\cal S}^a_{31,2}\rangle
\}.
\end{eqnarray}

The treating of such $\Psi$ is similar to the previous case, and we just list the final results as follows, 

\begin{eqnarray}
E &=&\frac {\langle \Psi |H| \Psi \rangle}{\langle \Psi|\Psi\rangle} \nonumber \\
&=& 0.632997 \lambda ^2  - 0.168735 \lambda  \nonumber \\
&=&    0.632997 (\lambda - 0.133283)^2 - 0.0112447.
\end{eqnarray}
Then the upper bound for ground-state energy is now improved to $-0.0112447$.

\section{Stability against electron-hole recombination}\label{stbl-eh-rcb}

The stability of a trion or a tetron against electron-hole recombination or annihilation actually is not a concern if the electrons and holes are from semi-metallic band-edge. Whether electrons ($e$) and holes ($h$) should recombine or not depends on the band structure. Here we just give a very simple example: semimetals. Electrons and holes coexist peacefully without annihilating one another, in semimetals \cite{semimetal} such as Bi, As, Sb. The reason is as follows and shown in Fig. 4. 

\begin{figure} [h]
\begin{center}
\leavevmode
\epsfxsize=3.1in 
\epsfysize=1.9in 
\epsffile{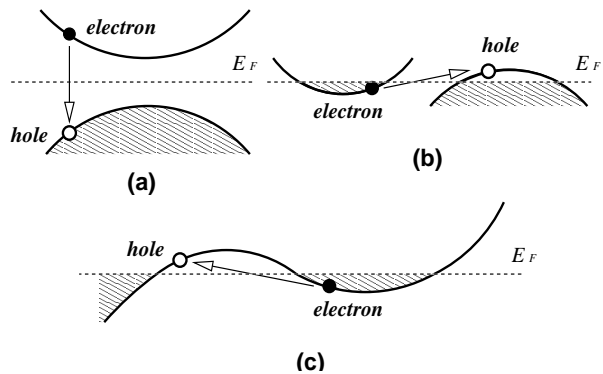}
\end{center}
\caption{Mechanism for the stability of electron-hole pairs. (a) Typical band structure for semiconductors. Electron states have higher energy than hole states. As a result, electrons recombine with holes in a short time. This reasoning also applies to metal provided the relevant conduction band is partly filled. (b) and (c) Band structures of semimetals. Electron states have lower energy than hole states and the recombination of electrons and holes is unfavored in energy. In all figures, $E_F$ is the Fermi energy.}
\label{fig4}      
\end{figure}

In typical semiconductors, the conduction and valence bands are separated by a positive energy gap as shown in Fig. 4(a). By jumping from the conduction to the valence band which is at a lower energy ({\it i.e.}, recombining with a hole), the thermally excited electron lowers its energy. This process thus occurs spontaneously to the $e$-$h$ pair and makes it unstable. (The same process can be interpreted as a hole jumping up and note that in contrast to electron, the higher is the hole in the band diagram, the lower is its energy.)  For semimetals with different band configurations, as typified in Fig. 4(b) and (c), the situation is however totally different. If the electron were to jump to the hole state and then recombine, as indicated by the arrows, it would have to take extra energy to reach there. This process is clearly unfavorable in energy and such a mechanism secures the stability of electron-hole liquid against recombination in semimetals. Thus we expect the proposed tetrons to find their best playing ground in solids with semi-metallic band structures.

\section{Stability against dissociation}\label{stbl-diso}

As mentioned before a tetron is not stable against the dissociation of one of its electrons into infinity in {\it vacuum}. To understand this, let us perform a thought experiment. The theorem by Lieb states that the ground-state energy of a 3E+1H complex must be higher than that of a trion plus an electron which is assumed to be infinitely far away, having arbitrarily small kinetic energy. Thus after we place a quasi-bound state of 3E+1H in vacuum, the complex tends to lower its energy by moving one of the electrons away to minimize the energy. This corresponds to the dissociation (detachment) of one electron as shown in Fig. 5. 

\begin{figure} [t]
\begin{center}
\epsfxsize=3.4in 
\epsfysize=1.8in 
\epsffile{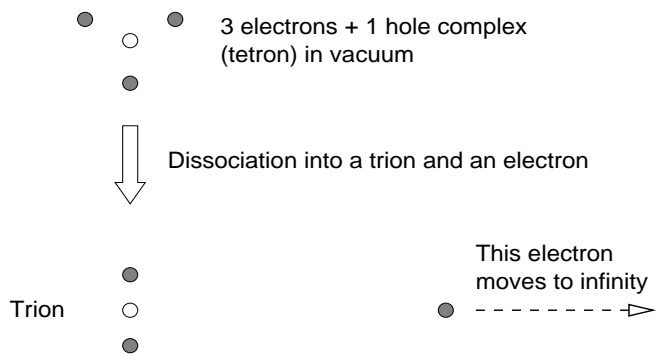}
\end{center}
\caption{The reason why tetron (3E+1H) is not stable in vacuum.}
\label{fig5}      
\end{figure}

However if the tetron is not in vacuum, for example, if many other electrons and holes are present nearby, then there is hardly any gain in energy for the whole system even if an electron dissociates from one tetron; this electron would simply increase Coulomb energy with other electrons anyway even if it decreases the energy of the original tetron as shown in Fig. 6. Such cases may happen to condensed matter systems where charge carrier densities are governed by electronic structure, or some types of external controls such as dopings. The electrons cannot leave the holes for infinity because statistically there must be some electrons close to the holes. When the densities of electrons and holes are low, the situation is more like a vacuum and therefore tetrons are not stable.  With an increasing charge density, however, the story could be increasingly different.  Then if the densities exceed a certain threshold, the whole system would lower its energy by taking advantage of the favorable geometric configuration offered by tetrons, provided the electrons outnumber holes by a ratio near or higher than 3. Then the tetron would be more stable in this high density electron-hole liquid. Actually Nature may have already suggested a similar mechanism for us: doubly charged negative ions until recently have not been seen in gas phase, or in isolation. In contrast, they have long been known to exist in condensed matter systems such as liquids and solutions\cite{science}. 

If the densities are to further increase by too much, the energy of the tetron state will no longer be negative after a second threshold. The reason is simple, the gain in Coulomb energy increases as $a^{-1}$ while the penalty paid for the kinetic energy scales as $a^{-2}$, where $a$ is the average distance between particles. Then we would see a system close to ordinary metal because the quasi-bound state is no longer possible. 

Here we just give the above qualitative argument to show how tetron states might be materialized. Further studies are needed in a more quantitative and rigorous way.

\begin{figure} [t]
\begin{center}
\epsfxsize=3.4in 
\epsfysize=2.9in 
\epsffile{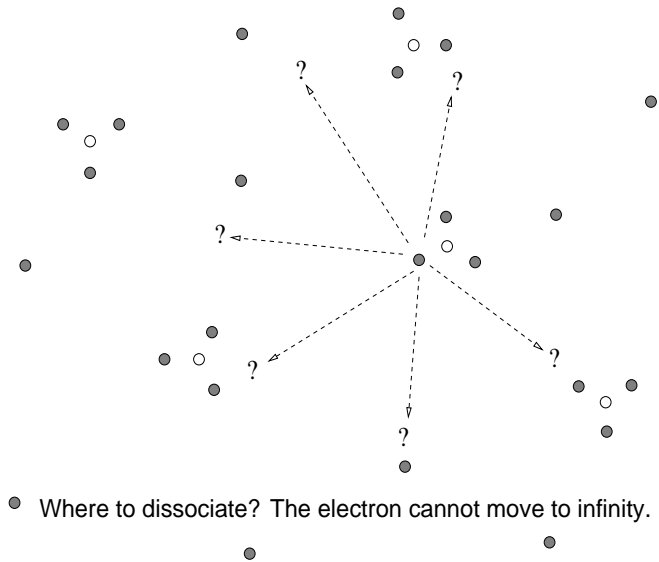}
\end{center}
\caption{The reason why tetron (3E+1H) could be stable in the electron-hole liquid system.}
\label{fig6}      
\end{figure} 

\section{Possible connection to high-$T_c$ pairing mechanism?}

The tetron state acts in a whole as a boson because it starts with 4 fermions. This is significant because the net charge is $-2e$ ($2e$ if we start with 3h+1e which is a more realistic scenario for most copper-oxide superconductors), and we are now left with a Cooper pair in effect. Moreover our current variational result implies a singlet state for the spin part of wavefunction, which is exactly what has been found in the Cooper pairs of high-$T_c$ superconductors\cite{burns}. This entices us into the possible connection between the tetron state and high-$T_c$ superconductivity, although this model itself could be oversimplified without considering the crystal structure of the host material. 

Charged fermion pairing models based on exchange of excitons have been proposed before by some notable theorists\cite{bardeen,ginzburg}, often in a paradigm of perturbative approaches. What we have developed here could be a non-perturbative way of treating similar problems.  

High-$T_c$ superconductivity found in copper oxides or cuprates are likely from the Bose-Einstein condensation of pre-formed pairs\cite{becondense}. In this regard, the tetron complex reasonably fits into such a picture, with its small size comparable to the coherence length of cuprate Cooper pairs. Also high-$T_c$ is intimately related to low dimensionality. Recall that the ground-state energy of a 2D exciton is four times as low as that in 3D, while $T_c$ decreases in cuprates when the superconductors are increasingly doped from a 2D conductor to 3D one.

There are mounting evidences showing the coexistence of electrons and holes in cuprate superconductors from key experiments such as angle-resolved photoemission spectroscopy\cite{arpes}  and transport properties\cite{NLuo,luo}. Finally the qualitative discussion given in Sec. \ref{stbl-diso} actually quite fits to the doping-dependent phase diagram of cuprate superconductors from underdoping to overdoping. Although the case at this moment is far from certain, the author sincerely hopes that this paper may generate some interests along this line of reasoning.

\section{Conclusion}

In conclusion, various trial functions for 3E+1H complex (tetron) were worked out by rigorous variational method to give an upper limit on the ground-state energy. The calculation suggests quasi-bound states for such a complex. The possible connection of such states to a pairing mechanism of high-$T_c$ is discussed.


\begin{references}

\bibitem{kheng} K.~Kheng {\it et al.}, Phys. Rev. Lett.~{\bf 71}, 1752 (1993).

\bibitem{huard} V.~Huard {\it et al.}, Phys. Rev. Lett.~{\bf 84}, 187 (2000). 
 
\bibitem{stebe1} B.~St\'{e}b\'{e} {\it et al.}, Phys. Rev. B~{\bf 56}, 12454 (1997).

\bibitem{thilagam} A.~Thilagam {\it et al.}, Phys. Rev. B~{\bf 555}, 7804 (1997).

\bibitem{bethe} H.A.~Bethe and  E.E.~Salpeter, {\it Quantum mechanics of one-and two-electron atoms} (Academic Press, New York, 1957), p. 154.

\bibitem{massey} H.S.W.~Massey, {\it Negative Ions} (Cambridge Univ. Press, London, 1976). 

\bibitem{acl} Y.~Belchenko, Rev. Sci. Instrum.~{\bf 64}, 1385 (1993).

\bibitem{fusion} A.J.T.~Holmes, Plasma Phys. Contr. Fus.~{\bf 34}, 653 (1992).

\bibitem{lieb1} E.H.~Lieb,  Phys. Rev. Lett.~{\bf 52}, 315 (1984).

\bibitem{lieb2} E.H.~Lieb,  Phys. Rev. B~{\bf 29}, 3018 (1984).

\bibitem{h2-0} F.~Robicheaux, R.P.~Wood and C.H.~Greene, Phys. Rev. B~{\bf 49}, 1866 (1994).

\bibitem{h2-1} T.~Sommerfeld, U.V.~Riss, H.D.~Meyer, L.S.~Cederbaum,  Phys. Rev. Lett.~{\bf 77}, 470 (1996).

\bibitem{h2-2}  T.~Sommerfeld, U.V.~Riss, H.D.~Meyer, L.S.~Cederbaum, Phys. Rev. A~{\bf 55}, 1903 (1997).

\bibitem{h2-3}  T.~Morishita, C.D.~Lin and C.G.~Bao,  Phys. Rev. Lett.~{\bf 80}, 464 (1998).

\bibitem{h2-4}  T.~Sommerfeld, Phys. Rev. Lett.~{\bf 85}, 956 (2000).

\bibitem{trion-more-stbl1} E.A.~Hylleraas, Phys. Rev. B~{\bf 71}, 491 (1947).

\bibitem{trion-more-stbl2} M.A.~Lampert, Phys. Rev. Lett.~{\bf 1}, 450 (1958).

\bibitem{semimetal} D.R.~Lovett, {\it Semimetals \& Narrow-bandgap Semiconductors} (Pion, London, 1977), p. 111.

\bibitem{science} M.K.~Scheller, R.N.~Compton and L.S.~Cederbaum, Science~{\bf 270}, 1160 (1995). 

\bibitem{bardeen} D.~Allender, J.~Bray and J.~Bardeen, Phys. Rev. B~{\bf 7}, 1020 (1973).

\bibitem{ginzburg} V.L.~Ginzburg, Usp. Fiz. Nauk~{\bf 101}, 185 (1970) [Sov. Phys.-Usp.~{\bf 13}, 335 (1970)]. 

\bibitem{becondense} See for example, Y.J.~Uemura {\em et al.}, Phys. Rev. Lett.~{\bf 62}, 2317 (1989).

\bibitem{burns} See for example, G.~Burns, {\it High-Temperature Superconductivity: an Introduction} (Academic Press, Boston, 1992); C.P.~Poole, Jr., H.A.~Farach and R.J.~Creswic, {\it Superconductivity} (Academic Press, San Diego, 1995); P.W.~Anderson, {\it The Theory of Superconductivity in the High $T_c$ Cuprates} (Princeton University Press, Princeton, 1997). 


\bibitem{arpes} Y.-D.~Chuang {\it et al.}, Phys. Rev. Lett.~{\bf 83}, 3717 (1999); D.L.~Feng {\it et al.}, preprint, cond-mat/9908056; A.D.~Gromko{\it et al.}, cond-mat/0003017. A.A.~Zakharov {\it et al.}, Phys. Rev. B~{\bf 61}, 115 (2000). T.~Yoshida {\it et al.}, preprint, cond-mat/0011172.

\bibitem{NLuo} N.~Luo, preprint, cond-mat/0003074.

\bibitem{luo} N.~Luo, accepted for publication by Physica~{\bf C}. 

\end{references}
\end{document}